\documentclass[aps, pre, twocolumn, superscriptaddress, reprint]{revtex4-1}
\setcitestyle{super}

\usepackage{graphicx}
\usepackage{bm}
\usepackage{dcolumn}
\usepackage{amssymb,amsmath}
\usepackage[colorlinks=true,allcolors=blue]{hyperref}
\usepackage{color}
\usepackage{siunitx}

\usepackage{physics}
\usepackage{lipsum}
\newcommand{\cOut}[1]{}
\usepackage[todonotes={textsize=tiny,textwidth=44pt},truncate=hyphenate,deletedmarkup=xout,addedmarkup=uwave]{changes}
\newcommand{\figref}[2]{\hyperref[#1]{\ref{#1}(#2)}}

\begin{document}

%\title{Experimental observation of the curvature-induced asymmetric spin-wave dispersion in hexagonal nanotubes}
\title{Symmetry- and curvature effects on spin waves in vortex-state hexagonal nanotubes}

\author{Lukas K\"orber}
\email{l.koerber@hzdr.de}
\affiliation{Helmholtz-Zentrum Dresden~-~Rossendorf, Institute of Ion Beam Physics and Materials Research, Bautzner Landstra{\ss}e 400, 01328 Dresden, Germany}
\affiliation{Fakultät Physik, Technische Universit\"at Dresden, D-01062 Dresden, Germany}

\author{Michael Zimmermann}
\affiliation{Fakultät für Physik, Universit\"at Regensburg, Universit\"atsstra{\ss}e 31, D-93053 Regensburg, Germany}

\author{Sebastian Wintz}
\affiliation{Swiss Light Source, Paul Scherrer Institut, 5232 Villigen PSI, Switzerland}
\affiliation{Max-Planck-Institut für Intelligente Systeme, 70569 Stuttgart, Germany}

\author{Simone Finizio}
\affiliation{Swiss Light Source, Paul Scherrer Institut, 5232 Villigen PSI, Switzerland}

\author{Matthias Kronseder}
\affiliation{Fakultät für Physik, Universit\"at Regensburg, Universit\"atsstra{\ss}e 31, D-93053 Regensburg, Germany}

\author{Dominique Bougeard}
\affiliation{Fakultät für Physik, Universit\"at Regensburg, Universit\"atsstra{\ss}e 31, D-93053 Regensburg, Germany}

\author{Florian Dirnberger}
\affiliation{Fakultät für Physik, Universit\"at Regensburg, Universit\"atsstra{\ss}e 31, D-93053 Regensburg, Germany}

\author{Markus Weigand}
\affiliation{Max-Planck-Institut für Intelligente Systeme, 70569 Stuttgart, Germany}
\affiliation{Helmholtz-Zentrum Berlin, 12489 Berlin, Germany}

\author{J\"org Raabe}
\affiliation{Swiss Light Source, Paul Scherrer Institut, 5232 Villigen PSI, Switzerland}

\author{Jorge A. Ot\'alora}
\affiliation{Departamento de Física, Universidad Católica del Norte, Avenida Angamos 0610, Casilla 1280, Antofagasta, Chile}
%\affiliation{Institute of Metallic Materials at the Leibniz Institute for Solid State and Materials Research, IFW, 01069 Dresden, Germany}
%\affiliation{Departamento de F\'isica, Universidad T\'ecnica Federico Santa Mar\'ia, Avenida Espa\~na 1680, Casilla 110-V, Valpara\'iso, Chile}

\author{Helmut Schultheiss}
\affiliation{Helmholtz-Zentrum Dresden~-~Rossendorf, Institute of Ion Beam Physics and Materials Research, Bautzner Landstra{\ss}e 400, 01328 Dresden, Germany}
\affiliation{Fakultät Physik, Technische Universit\"at Dresden, D-01062 Dresden, Germany}
%\affiliation{Technische Universit\"at Dresden, D-01062 Dresden, Germany}

\author{Elisabeth Josten}
\affiliation{Ernst Ruska-Centre for Microscopy and Spectroscopy with Electrons (ER-C) and Peter Gr\"unberg Institute (PGI), Forschungszentrum J\"ulich, 52425 J\"ulich, Germany}

\author{J\"urgen Lindner}
\affiliation{Helmholtz-Zentrum Dresden~-~Rossendorf, Institute of Ion Beam Physics and Materials Research, Bautzner Landstra{\ss}e 400, 01328 Dresden, Germany}

\author{Istv\'an K\'ezsm\'arki}
\affiliation{Experimental Physics V, University of Augsburg, 86135 Augsburg, Germany}
%\author{J\"urgen Fassbender}
%\affiliation{Helmholtz-Zentrum Dresden~-~Rossendorf, Institute of Ion Beam Physics and Materials Research, Bautzner Landstra{\ss}e 400, 01328 Dresden, Germany}

\author{Christian H. Back}
\affiliation{Physik-Department, Technische Universit\"at M\"unchen, 85748 Garching b. M\"unchen, Germany}
\affiliation{Fakultät für Physik, Universit\"at Regensburg, Universit\"atsstra{\ss}e 31, D-93053 Regensburg, Germany}

\author{Attila K\'akay}
\affiliation{Helmholtz-Zentrum Dresden~-~Rossendorf, Institute of Ion Beam Physics and Materials Research, Bautzner Landstra{\ss}e 400, 01328 Dresden, Germany}

%\abbreviations{DMI, Dzyaloshinskii–Moriya interaction; FFT, fast Fourier transform; SW, spin wave; TR-STXM, time-resolved scanning transmission X-ray microscopy}

%\newpage
\begin{abstract}

Analytic and numerical studies on curved magnetic nano-objects predict numerous exciting effects that can be referred to as magneto-chiral effects, which do not originate from intrinsic Dzyaloshinskii–Moriya interaction or interface-induced anisotropies. In constrast, these chiral effects stem from isotropic exchange or dipole-dipole interaction, present in all magnetic materials, which acquire asymmetric contributions in case of curved geometry of the specimen. As a result, for example, the spin-wave dispersion in round magnetic nanotubes becomes asymmetric, namely spin waves of the same frequency propagating in opposite directions along the nanotube exhibit different wavelenghts. Here, using time-resolved scanning transmission X-ray microscopy experiments, standard micromagntic simulations and a dynamic-matrix approach, we show that the spin-wave spectrum undergoes additional drastic changes when transitioning from a continuous to a discrete rotational symmetry, \textit{i.e.} from round to hexagonal nanotubes, which are much easier to fabricate. The polygonal shape introduces  localization  of  the  modes  both to  the  sharp, highly curved corners  and  flat  edges.   Moreover, due to the discrete rotational symmetry, the  degenerate  nature  of  the  modes  with  azimuthal  wave  vectors  known from round tubes is partly lifted, resulting in singlet and duplet modes. For comparison with our experiments, we calculate the microwave absorption from the numerically obtained mode profiles which shows that a dedicated antenna design is paramount for magnonic applications in 3D nano-structures. To our knowledge these are the first experiments directly showing real space spin-wave propagation in 3D nano objects.

\end{abstract}

\maketitle

\section{Introduction}

After having been proposed by Bloch in the 1930s,\cite{blochZurTheorieFerromagnetismus1930} the propagation of spin waves (SWs) -- the elementary excitations in magnetically ordered systems -- has been studied extensively in the past. Because of their peculiar linear and nonlinear characteristics, SWs promise great potential in information transport and processing as, \textit{e.g.}, the magnon transistor \cite{chumakMagnonTransistorAllmagnon2014} and the magnonic diode \cite{lanSpinWaveDiode2015} for multifunctional spin-wave logic applications. Spin waves (including the spatially uniform ferromagnetic resonance precession) have also been proven to be an excellent tool to probe the magnetic characteristics of solids as they are sensitive to spin currents,\cite{vlaminckCurrentInducedSpinWaveDoppler2008,chauleauSelfconsistentDeterminationKey2014,gladiiSpinwavePropagationSpinpolarized2017}
%"Self-consistent determination of the key spin-transfer torque parameters from SW Doppler experiments" J.-Y. Chauleau, H.G. Bauer, H.S. Koerner, J. Stigloher, M. Haertinger, G. Woltersdorf, C.H. Back, Phys. Rev. B 89, 020403(R) (2014).
impurities,\cite{callawayScatteringSpinWaves1964,abeedEffectMaterialDefects2019,mohseniBackscatteringImmunityDipoleexchange2019} crystal anisotropies \cite{gurevichMagnetizationOscillationsWaves1996} or asymmetric exchange interactions, among others. For example, the presence of an asymmetric interaction such as the Dzyaloshinskii–Moriya interaction (DMI) leads to an asymmetric dispersion and consequently to a nonreciprocal propagation of spin waves, therein.\cite{udvardiChiralAsymmetrySpinwave2009,zakeriAsymmetricSpinWaveDispersion2010,kezsmarkiEnhancedDirectionalDichroism2011a,bordacsChiralityMatterShows2012a,szallerSymmetryConditionsNonreciprocal2013,moonSpinwavePropagationPresence2013,cortes-ortunoInfluenceDzyaloshinskiiMoriya2013,kostylevInterfaceBoundaryConditions2014,kezsmarkiOnewayTransparencyFourcoloured2014a,kornerInterfacialDzyaloshinskiiMoriyaInteraction2015,kezsmarkiOpticalDiodeEffect2015} Similar non-reciprocal spin-wave propagation is observed in magnetic bilayers. \cite{grunbergMagnetostaticSpinWave1981,henryPropagatingSpinwaveNormal2016,gallardoReconfigurableSpinWaveNonreciprocity2019}
Therefore, the study of spin-wave propagation is both of a technological as well as a fundamental interest. 

While many of the aforementioned effects have been investigated mostly in bulk or in flat thin-film samples, over the last decade, curvature-induced effects have been uncovered as a new way to manipulate magnetic equilibria and spin dynamics. Numerous analytic and numerical works have already shown that the surface curvature and geometry of three-dimensional magnetic membranes leads to phenomena not present in flat specimen of the same material.\cite{landerosReversalModesMagnetic2007,yanChiralSymmetryBreaking2012,otaloraBreakingChiralSymmetry2013,yanSpinCherenkovEffectMagnonic2013,kravchukCurvatureEffectsThin2014,pylypovskyiGeometryinducedEffectsDomain2015,otaloraCurvatureInducedAsymmetricSpinWave2016,otaloraAsymmetricSpinwaveDispersion2017,otaloraFrequencyLinewidthDecay2018,kravchukMultipletSkyrmionStates2018} For example, in conventional soft magnetic materials, exotic non-collinear magnetic textures such as skyrmions\cite{kravchukTopologicallyStableMagnetization2016} may be stabilized by bending the magnetic material.  Moreover, magnetization dynamics can be influenced, leading to symmetry breaking of domain-wall motion,\cite{yanChiralSymmetryBreaking2012,landerosReversalModesMagnetic2007} asymmetric spin-wave transport\cite{hertelCurvatureInducedMagnetochirality2013,otaloraCurvatureInducedAsymmetricSpinWave2016} or the emergence of a topological Berry phase.\cite{dugaevBerryPhaseMagnons2005} 
The influence of curvature on magnetic equilibria has been shown to be mainly due to a renormalization of the magnetic exchange interaction. \cite{kravchukCurvatureEffectsThin2014,shekaNonlocalChiralSymmetry2020} However, as shown in Ref.~\citenum{shekaNonlocalChiralSymmetry2020}, the long range dipole-dipole interaction can also lead to chiral symmetry breaking effects and thus introducing handedness in an intrinsically achiral material. 
%On one hand, the magnetization dynamics in perturbed by curvature-induced magnetic charges,\cite{landerosReversalModesMagnetic2007,yanChiralSymmetryBreaking2012,yanSpinCherenkovEffectMagnonic2013} \textit{i.e.} by a renormalization of the dipolar interaction. 
In our recent works Refs.~\citenum{otaloraCurvatureInducedAsymmetricSpinWave2016,otaloraAsymmetricSpinwaveDispersion2017,salazar-cardonaNonreciprocitySpinWaves2021}, we predicted that curvature-induced effects, both of dipolar and exchange origin, lead to an asymmetric dispersion of SWs in round magnetic nanotubes being in the helical state. 
%On the other hand, the curvature-induced effect of exchange origin is responsible for the emergence of a topological Berry phase predicted for rings by Dugaev \textit{et al.}\cite{dugaevBerryPhaseMagnons2005} and recently shown for tubes by Otálora \textit{et al.}\cite{salazar-cardonaNonreciprocitySpinWaves2021}

In this manuscript we show that magnetization in 3D nano-objects is not only governed by the curvature and topology. At the example of hexagonal nanotubes we present that the discrete rotational symmetry induces drastic changes in the spin-wave spectra. In particular, the mode spectrum is non-trivially split into standing waves (singlets) and running waves (duplets). %This is a quality specifically not present in our previously investigated round nanotubes.
In the previously investigated round nanotubes the modes with the same $k$ but opposite sign of azimuthal quantization index form duplets, except for the uniform mode, which is a singlet. Moreover, in contrast to round magnetic nanotubes, which are extremely difficult to fabricate with sufficiently low damping, we succeeded to fabricate hexagonal nanotubes and prepare them in the vortex magnetic state.\cite{zimmermannOriginManipulationStable2018} This allowed us, for the first time, to directly image magnetization dynamics in curved 3D nanostructures, using time-resolved scanning transmission X-ray microscopy (TR-STXM).\cite{vanwaeyenbergeMagneticVortexCore2006,acremannSoftwareDefinedPhoton2007,wintzMagneticVortexCores2016} Prior to this work, experimental evidence in curved samples was restricted to equilibrium effects, such as the spiral Landau pattern in bent rectangular elements~\cite{dietrichInfluencePerpendicularMagnetic2008} or the domain wall pinning by a curvature gradient, shown in parabolic stripes.\cite{volkovExperimentalObservationExchangeDriven2019}

For the quantitative analysis of the experiments, we have conducted an extensive numerical study using standard micromagnetic simulations as well as our recently developed finite-element propagating-wave dynamic-matrix approach.\cite{korberFiniteelementDynamicmatrixApproach2021}  We find that the mode spectrum of vortex-state hexagonal tubes is asymmetric and quite complex in nature. The polygonal shape introduces localization of the modes into the highly curved corners and flat facets. Moreover, the degenerate nature of the modes with azimuthal wave vectors known from round tubes is lifted in the polygonal case, resulting in singlet and duplet modes. The singlet-duplet differentiation is related to the discrete symmetry resulting from the hexagonal cross section of the waveguide housing a vortex magnetic ground state. Using the spin-wave profiles resulting from our eigensolver, we calculated the dispersion relation with the microwave absorption for two different antenna field profiles, namely for a current loop and a stripline antenna. The numerical results show that the stripline antenna used in the TR-STXM experiments excites multiple modes at a fixed frequency, though with different intensity, thus the resulting spin waves propagating in the nanotube form a beating pattern instead of a single wave with a well defined wavelength. Therefore to measure and eventually exploit the asymmetry of the dispersion a stripline antenna is not satisfactory. Instead, a proper design of the microwave antenna is needed to couple and thus excite only single modes with well defined frequency and wavelength.

%\textbf{\color{red}TODO LK: Introduction has to be way shorter, method section can't start at the end of page 2.}
In Sec.~\ref{sec:methods}, we briefly discuss the sample fabrication followed by the description of the micromagnetic methods used in the manuscript, employed both in the time domain and in the frequency domain. The experimentally measured spin-wave propagation together with finite-element micromagnetic simulations is shown in Sec.~\ref{sec:results}. To understand the spin-wave propagation in hexagonal tubes the dispersion relation is discussed in detail, showing the localization of modes and their singlet-duplet nature related to the combined symmetry of the tube geometry and its ground magnetic state. The predicted microwave absorption and its effect on the spin-wave excitation depending on the antenna geometry is discussed to allow for a realistic comparison of the numerical and experimental results. The conclusions of the study and a possible outlook, with suggestions for further experimental investigations, will be discussed lastly in Sec.~\ref{sec:outlook}.

\section{Methods}\label{sec:methods}

In this section we will shortly summarize the numerical as well as the experimental methods, including the sample fabrication involved in the current study.

\subsection{Sample fabrication and STXM experiments}

The nanotube fabrication involves a two step routine: First, GaAs rods are grown on oxidized Si(111) wafers via molecular-beam epitaxy (MBE) in a III-V MBE using Ga droplets as catalysts. After in situ transfer to a metallic MBE chamber in a pressure lower than $1\times 10^{-10}$ mbar, the coating layers composed of the permalloy magnetic layer and an Al capping layer to avoid oxidation are deposited at pressures around $1\times 10^{-10}$ mbar (base pressure of $5\times 10^{-11}$ mbar). Details of the sample preparation are described in Ref.~ \citenum{zimmermannOriginManipulationStable2018}. As a result of the growth-induced easy-plane magnetic anisotropy perpendicular to the symmetry axis, the equilibrium magnetization is a vortex state, as confirmed by STXM measurements.\cite{zimmermannOriginManipulationStable2018}

\begin{figure}[h!]
\begin{center}
\includegraphics[width=8.6cm]{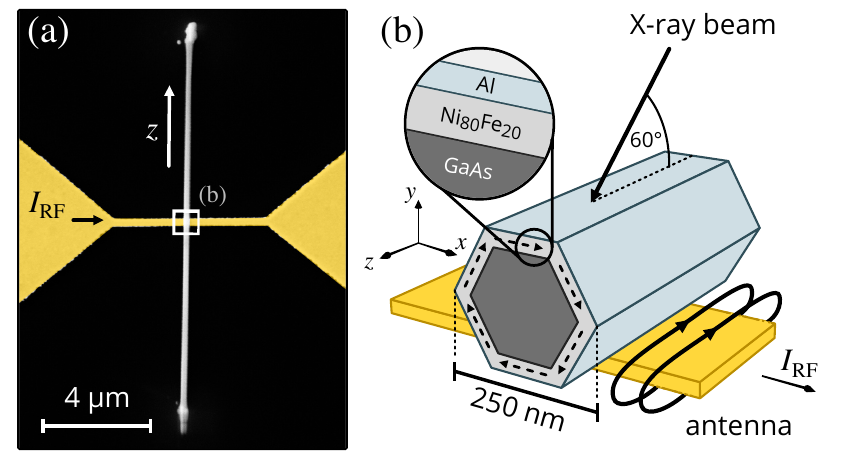}
\caption{\label{fig:fig1}(a) Scanning-electron-transmission-microscopy image of a hexagonal permalloy nanotube with \SI{250}{\nano\meter}  outer diameter, \SI{30}{\nano\meter} thickness and \SI{12}{\micro\meter} length on a GaAs wire. The gold stripline antenna (here, colored for visual purposes) was patterned on a SiN membrane and the nanotube was placed on the top using a focused ion beam (FIB) tool and a micro manipulator. The Oersted field of an rf-current is used to excite SWs. (b) Cross-sectional sketch of the hexagonal nanotube showing the layer structure. The permalloy layer is directly evaporated on the GaAs wire \textit{in situ} and capped with Al to avoid oxidation. A relative angle of 60 degree was used between the X-ray beam and the symmetry axis of the nanotube. This configuration allows for being sensitive to the in-plane dynamic magnetization of the top- and bottom surfaces of the tube, which is expected to be larger than the out-of-plane dynamic magnetization component.}
\end{center}
\end{figure}

The TR-STXM measurements were performed to directly image the magnetization dynamics in the tubes in hope for  experimentally obtain the spin-wave dispersion. The time-resolved measurements were mostly performed at the MAXYMUS endstation of BESSY II at Helmholtz-Zentrum Berlin, Germany. The static and low-frequency (up to 6 GHz) measurements have been done at the POLLUX endstation of the PSI, Villigen, Switzerland.

An exemplary scanning-electron-transmission-microscopy image of a \SI{250}{\nano\meter} outer diameter nanotube  used for the measurements is shown in Figure~\ref{fig:fig1} together with a sketch of the nanotube and its cross-sectional view. The magnetic nanotube is placed on the top of a gold stripline antenna patterned on a SiN membrane. The nanotube is usually oriented such that one of the flat facets is parallel to the substrate surface thus the top facet normal is parallel to the X-ray beam. The absorption spectra were collected by exploiting the X-ray magnetic circular dichroism (XMCD) effect~\cite{schtutzAbsorptionCircularlyPolarized1987} of the transmitted soft X-ray radiation at the L3-edge of iron (708 eV). The dynamic magnetization contrast to visualize the real space spin-wave propagation was obtained with left circularly polarized light. A Fresnel zone plate is used to focus the X-rays to a single spot on the sample, allowing for a lateral resolution of approximately \SI{25}{\nano\meter} when the sample is raster scanned through the beam. The acquired magnetic contrast scales with the projection of the magnetic orientation on the direction of photon propagation. Hence, in normal incidence, our STXM set-up is sensitive to the dynamic magnetization component in the propagation direction perpendicular to the top and bottom surfaces. While the 30 degree inclined sample mounting, compared to the surface normal, used in the experiments also allows for detecting in-plane magnetization components at the same time. The spin waves were excited with the stripline antenna at various frequencies between a frequencies of \SI{1}{GHz} and \SI{10}{GHz}. Every excitation yielded a 7 frame movie; each frame contains the real-space profiles of the excited spin waves at equidistant phases with respect to the excitation signal.

\subsection{Micromagnetic modeling}

In this Section we introduce the micromagnetic methods used to investigate the spin-wave propagation and dispersion in our hexagonal nanotubes.

\subsubsection{Finite-element time-domain simulations}
In the framework of micromagnetism the magnetization dynamics is described by the Landau-Lifshitz-Gilbert equation of motion, 

\begin{equation}\label{eq:llg}
    \frac{\mathrm{d}\bm{m}}{\mathrm{d}t} = -\omega_M (\bm{m} \times  \bm{h}_\mathrm{eff}) + \alpha_\mathrm{G} \left(\bm{m}\times \frac{\mathrm{d}\bm{m}}{\mathrm{d}t}\right)
\end{equation}
where $\bm{m}$ is the reduced magnetization $\bm{m}(\bm{r},t) = \bm{M}(\bm{r},t)/M_\mathrm{s}$, $M_\mathrm{s}$ the saturation magnetization, $\bm{h}_\mathrm{eff}$ the normalized effective field, $\omega_M = \gamma \mu_0 M_\mathrm{s}$ the characteristic frequency and $\alpha_\mathrm{G}$ the Gilbert damping parameter. In order to study the propagation of the spin waves in the hexagonal tubes we have solved numerically the equation of motion using our custom developed GPU accelerated finite-element micromagnetic code \textsc{TetraMag}.\cite{kakaySpeedupFEMMicromagnetic2010} For the simulations we have considered an \SI{8}{\micro\meter} long hexagonal tube with \SI{30}{\nano\meter} thickness assuming permalloy material parameters. The exact values can be seen in Table~\ref{tab:matparam}. The equilibrium state, which is a flux-closure state is computed using a conjugate-gradient energy minimization starting from a circular vortex initial state. The uniaxial anisotropy along the long axis of the tube with a negative constant will prefer a flux-closure (vortex) state. In order to mimique the experimental excitation scheme, monochromatic spin waves were excited at the center of the hexagonal tube using a microwave field of \SI{1.0}{\milli\tesla} magnitude, with components in the $yz$ plane, and with the spatial profile of the field corresponding to the stripline antenna of \SI{250}{\nano\meter}. The rf field with a sinusoidal time variation was applied for 100 periods for all simulated frequencies.

\begin{table}[h!]
\caption{\label{tab:matparam}Parameters used for micromagnetic modeling. 
%\textbf{TODO Attila these parameters are still not right}
}
\begin{ruledtabular}
\begin{tabular}{ll}
		exchange stiffness ($A_\mathrm{ex}$) & \SI{13}{\pico\joule/\meter}\\
		saturation ($M_\mathrm{s}$) & \SI{820}{\kilo\ampere/m}\\
		reduced gyromagnetic ratio ($\gamma/2\pi$) & \SI{28}{\giga\hertz/\tesla}\\
		Gilbert damping ($\alpha_\mathrm{G}$) & 0.007\\
		uniaxial anisotropy constant ($K_{\mathrm{u},1}$) & \SI{-20}{\kilo\joule\per\cubic\meter} \\
		uniaxial anisotropy direction ($\bm{e}_{\mathrm{u}}$) & $\bm{e}_z$ \\
		tube outer diameter ($D$) & \SI{250}{\nano\meter}\\
		tube shell thickness ($T$) & \SI{30}{\nano\meter}\\
		tube length, $(L)$ & \SI{8,0}{\micro\meter}\\
			edge length along tube& \SI{5}{\nano\meter}\\
		edge length along cross-section & \SI{3}{\nano\meter}
\end{tabular}
\end{ruledtabular}
\end{table}

\subsubsection{Propagating-wave dynamic-matrix approach}\label{sec:dynmat}

To numerically calculate the spin-wave dispersion for waves travelling along the hexagonal nanotube we utilize our recently developed finite-element dynamic-approach for propagating waves. This approach uses the same spatial discretization method as \textsc{TetraMag}\cite{kakaySpeedupFEMMicromagnetic2010} and relies on the numerical solution of the eigenvalue problem 
\begin{equation}\label{eq:linllg}
    \frac{\omega_\nu(k)}{\omega_M} \bm{\eta}_{\nu k} = i\bm{m}_0 \times \hat{\mathbf{\Omega}}_k \bm{\eta}_{\nu k} \quad\text{with}\quad \bm{m}_0 \perp \bm{\eta}_{\nu k}
\end{equation}
which is the non-dissipative ($\alpha_\mathrm{G}=0$) version of the LLG equation \eqref{eq:llg}, linearized in the vicinity of some (stable) equilibrium state $\bm{m}_0(\bm{r})$ and transformed into a single cross section of the nanotube for the case of plane waves propagating along the $z$ direction with wave vector $k$ and angular frequency $\omega_\nu (k)$. Note, the $z$ axis in this study is the axis along the long axis of the hexagonal nanotube. The eigenvectors $\bm{\eta}_{\nu k} \equiv \bm{\eta}_{\nu k}(\bm{\rho})$ represent the (complex) lateral mode profiles which only depend on the coordinates $\bm{\rho} = (x,y)^T$ and can be denoted additionally by some lateral mode index $\nu$ which labels the respective branch of the dispersion.

The plane-wave Hamiltonian operator $\hat{\mathbf{\Omega}}_k$ is given by
\begin{equation}
    \hat{\mathbf{\Omega}}_k = h_0 \hat{\mathbf{I}} + \hat{\mathbf{N}}_k = h_0 \hat{\mathbf{I}} + e^{-ikz}\hat{\mathbf{N}}e^{ikz}.
\end{equation}
with $h_0$ being the projection of the unitless static effective field (including any static external field) onto to equilibrium direction, $\Hat{\mathbf{I}}$ the identity operator and $\Hat{\mathbf{N}}$ the self-adjoint operator describing the magnetic self interactions, which, in our case, are exchange and dipolar interaction, as well as uniaxial magnetic anisotropy (see Ref.~\citenum{korberFiniteelementDynamicmatrixApproach2021}). All operators and vectors are discretized in the framework of the finite-element method. The resulting linear system is numerically diagonalized for each $k$ value using the iterative Arnoldi-L\'anczos method which yields a desired number of lowest-magnitude eigenvalues $\omega_\nu(k)$ as well as their corresponding eigenvectors $\bm{\eta}_{\nu k}$. In our case, the lowest 30 modes for each $k$ were calculated for a total number of 201 wave vectors between \SI{\pm35}{\radian/\micro\meter}. To account for the dipolar potential generated by the individual spin-wave modes we employ a modified version of the hybrid FEM/BEM Fredkin-Koehler method which was recently extended to plane-wave potentials in Ref.~\citenum{korberFiniteelementDynamicmatrixApproach2021}. The equilibrium state $\bm{m}_0$ is found by energy minimization, the same way as for the time-domain simulations.

In contrast to a full 3D time-domain simulation, the magnetic nanotube only needs to be modeled in a single cross section which drastically reduces the computational load. The spin-wave frequencies and mode profiles are directly obtained (within minutes), without the need of additional post processing. Moreover, as an additional benefit, degenerate modes can be detected which is not easily possible using a single field pulse followed by an FFT-based analysis. 

\section{Results and Discussion}\label{sec:results}

In this section we will first show the spin-wave propagation measured experimentally using time-resolved STXM and compare it with those from micromagnetic simulations. We would like to emphasize that according to our knowledge these are the first experiments directly showing real space spin-wave propagation in 3D nano objects.

\subsection{Mode localization}

To obtain a first overview of the spin-wave transport in hexagonal nanotubes we excite monochromatic waves using a stripline microwave antenna at the center of the tube and at different fixed frequencies. 

\begin{figure}[h!]
\begin{center}
\includegraphics[width=\columnwidth]{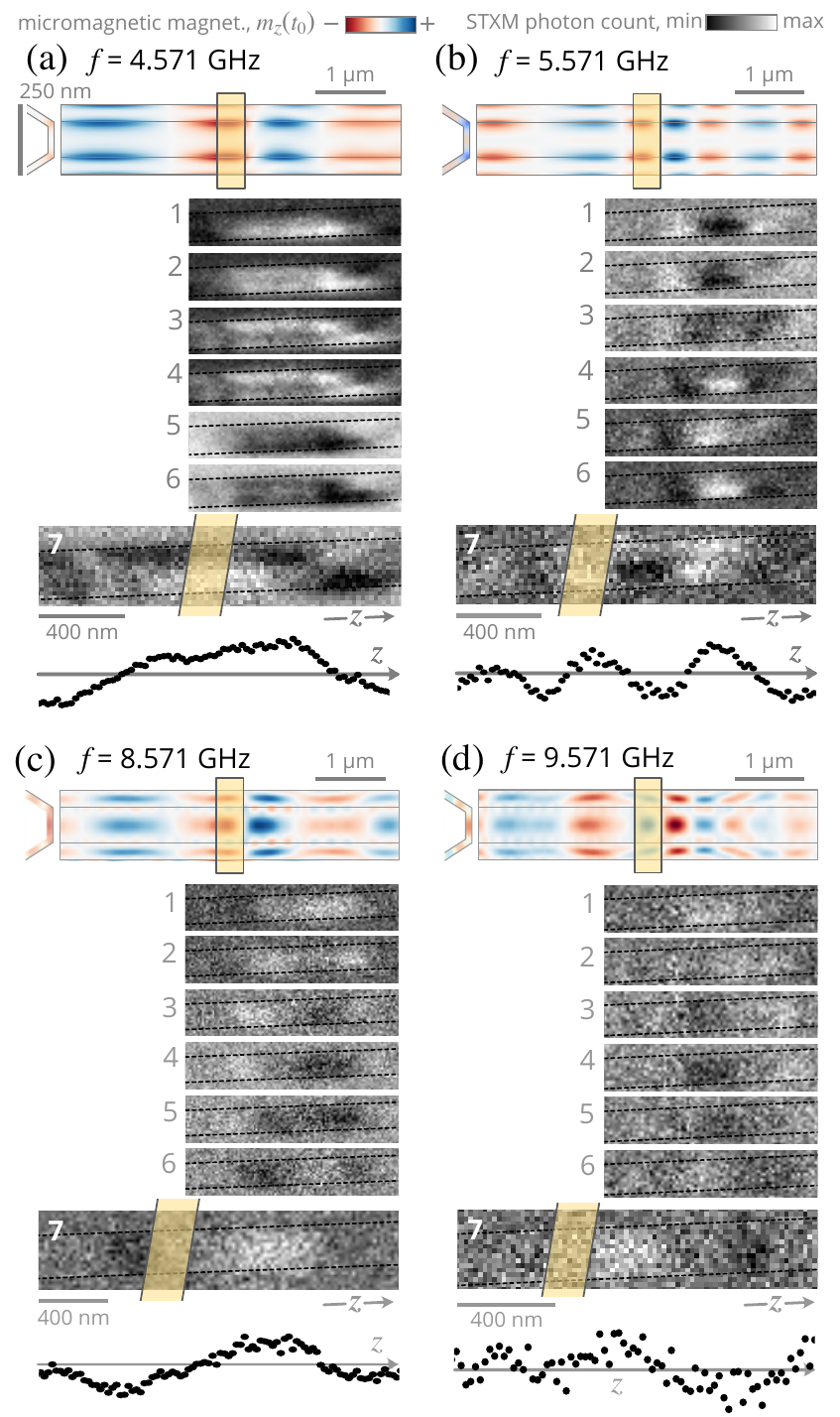}
\caption{\label{fig:fig2}{Numerical (red-blue) and experimental (gray scale) snapshots of the dynamic magnetization at fixed times for modes propagating in the corners of the hexagonal tube (a) at \SI{5.571}{\giga\hertz} and (b) at \SI{4.571}{\giga\hertz}, and for modes mostly propagating in the top and bottom facets (c) at \SI{8.571}{\giga\hertz} and (d) at \SI{9.571}{\giga\hertz}. For the numerical profiles, a side view is shown next to the projection of the upper half the nanotube. For better illustration, they have been stretched in the width direction. Below we show all frames of the respective STXM movies. In addition to each last frame, we show an average linescans along the tube, together with sinusoidal fits used to obtain the wave lengths. The position of the antenna is in all cases marked with a translucent gold-colored patch.}}
\end{center}
\end{figure}

In Fig.~\figref{fig:fig2}{a-d}, we show for exemplary snapshots of two counter-propagating spin-wave modes at \SI{4.571}{\giga\hertz}, \SI{5.571}{\giga\hertz}, \SI{8.571}{\giga\hertz} and \SI{9.571}{\giga\hertz}, obtained by TR-STXM and time-domain micromagnetic simulation. For the experiments, we show a full oscillation cycle as seven frames. As seen especially from the simulation profiles, the modes exhibit different localization within the cross section of the hexagonal tube, \textit{e.g.}, there are modes more localized in the corners of the tube (\SI{5.571}{\giga\hertz} in Fig.~\figref{fig:fig2}{a} and \SI{4.571}{\giga\hertz} in Fig.~\figref{fig:fig2}{b}) or on the facets of the tube (\SI{8.571}{\giga\hertz} and \SI{9.571}{\giga\hertz} in Fig.~\figref{fig:fig2}{c,d}). We also would like to refer to the {animated} experimental movies of these modes, provided in the supplemental material, which {may show the localization of the modes at \SI{4.571}{\giga\hertz} and \SI{5.571}{\giga\hertz} in the corners better than the static frames}. We observe an intensity asymmetry of the modes at large frequencies. This is a commonly known effect for Damon-Eshbach SWs (with $\bm{k}\perp\bm{m}_0$) excited with a stripline antenna in magnetic thin films,\cite{demidovExcitationMicrowaveguideModes2009} and is also present here as our tubes are in the vortex state ($\bm{k}\perp\bm{m}_0$). Moreover, in the numerical mode {snapshots at \SI{5.571}{\giga\hertz} (Fig.~\figref{fig:fig2}{a}) and \SI{4.571}{\giga\hertz} (Fig.~\figref{fig:fig2}{b})}, one can already clearly see a wave-vector asymmetry for the two counter-propagating modes which is the evidence for an asymmetric SW dispersion. 

To obtain the wavelenghts of the counter-propagating spin waves from the experimental data, we average the measured data for different excitation frequencies along the width of the tube (shown for each last STXM frame in Fig.~\ref{fig:fig2}) and fit these curves for individual time frames with decaying sinusoidal functions along the long axis of the tube (propagation direction) on either side of the antenna. Between 2 and 18 fits could be obtained per frequency and direction. The resulting wave vectors were obtained as the averages of all associated fits. This method was chosen because only a few wavelengths are observed in the measurements, and a Fourier analysis to obtain the wave vectors at a given frequency was not conclusive. {Note, that from the two-sided fit presented for the corner mode at \SI{5.571}{\giga\hertz} in Fig.~\figref{fig:fig2}{a}, one can also see a wave-vector asymmetry in the experimental data}

\subsection{Dispersion and mode symmetry}

\begin{figure}[h!]
    \centering
    \includegraphics[width=8.6cm]{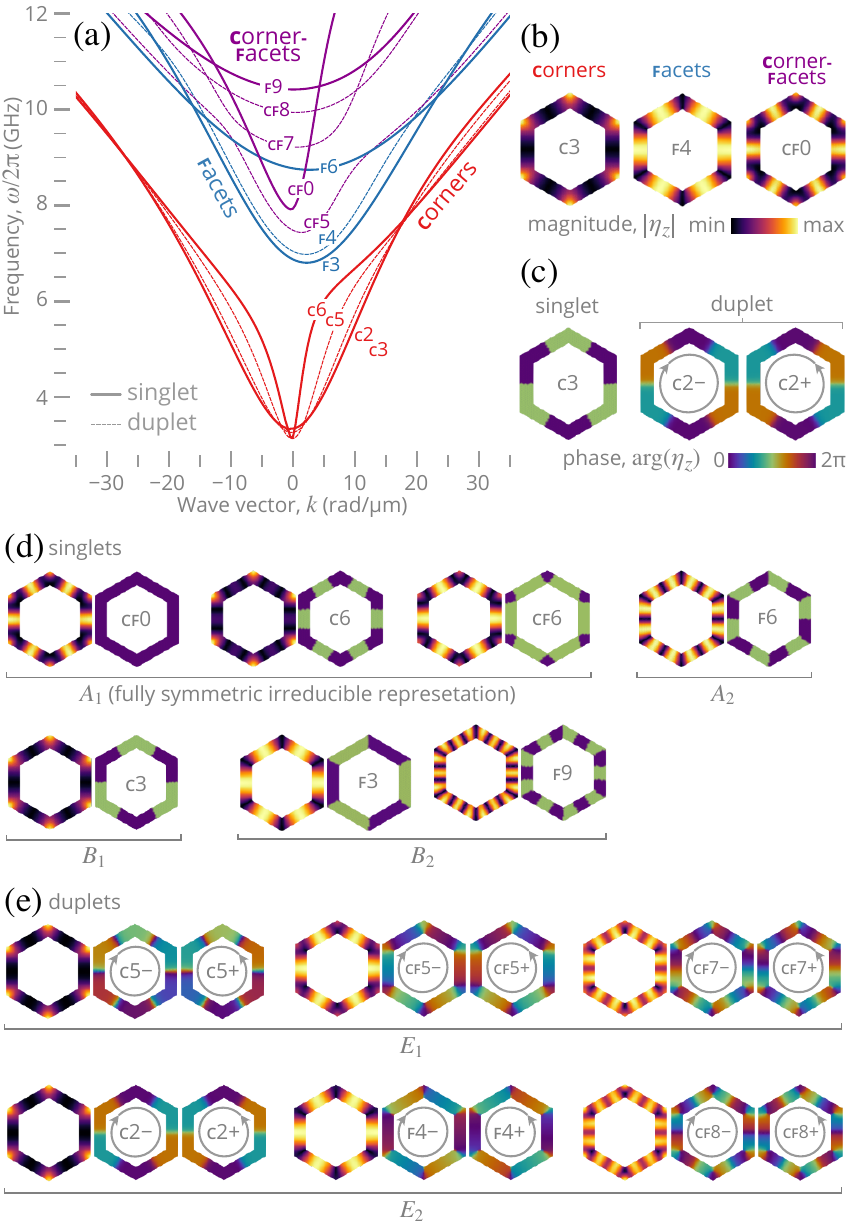}
    \caption{Dispersion relation and modes profiles directly resulting from the dynamic-matrix approach for propagating spin-waves. (a) Full dispersion for all modes below \SI{12}{\giga\hertz}. The modes can be categorized based on their localization position as corner \textsc{c}, facet \textsc{f} and corner-facet \textsc{cf} modes, shown in (b). A second categorization can be made based on their singlet and duplet nature, by looking at the phase of the wave along the azimuthal direction in (c); singlets are standing spin-wave solutions while duplets are two degenerate solutions of counter propagating waves along the azimuthal direction. In panels (d) and (e) the mode magnitude as well as its phase is shown for several examples of the singlet and duplet solutions. Below the modes, we annotate the corresponding irreducible representations which they belong to.}
    \label{fig:fulldisp} 
\end{figure}

For a more detailed analysis of the modal spectrum, we calculate the full dispersion for all modes below \SI{12}{\giga\hertz} and with wave vectors between \SI{\pm35}{\radian/\micro\meter} using the FEM propagating-wave dynamic matrix approach outlined in Sec.~\ref{sec:methods} and explained in detail in Ref.~\citenum{korberFiniteelementDynamicmatrixApproach2021}. We would like to point out again that, in contrast to usual time-domain simulations based on microwave excitation, this analysis provides access to modes which might have a nontrivial spatial profile and which therefore do not couple to commonly used microwave-field distributions. Moreover, it becomes possible to always separate degenerate modes. The resulting dispersion including all modes is shown in Fig.~\figref{fig:fulldisp}{a}. As suggested by the varying mode localization at different excitation frequencies in the previous section, the spectrum is divided into several, in certain cases, even degenerate branches. We will see in the next section, that only a few of them are easily accessible in experiments. The different branches $\nu$ can be categorized by analyzing the corresponding (complex-valued) lateral mode profiles $\bm{\eta}_{\nu k}(\bm{\rho})$ within the hexagonal-tube cross section. To visualize these, we plot the magnitude $\abs{\eta_z}$ and the phase $\mathrm{arg}(\eta_z)$ of the $z$ component of the mode profiles as color maps in Fig.~\ref{fig:fulldisp}. 

As indicated by the magnitudes of the spatial profiles in Fig.~\figref{fig:fulldisp}{b}, the branches can be categorized by their localization either to the corners (\textsc{c}) or to the facets (\textsc{f}) of the hexagonal tube. Moreover, there are also several hybrid-corner-facet modes (\textsc{cf}). Apart from their localization, the modes differ in the number of periods along the hexagonal circumference. In this sense, the spin waves in a hexagonal tube are similar to the ones in round nanotubes or rings in the vortex state where the azimuthal dependence of the mode profiles is given by $\exp(im\phi)$ with $m$ being an integer number often called the azimuthal mode index. In such cylindrical or tubular systems, modes with the same $k$ but opposite sign of $m$ are degenerate, i.e. they form duplets, except for the m=0 mode, which is a singlet. In the phase plots such as Fig.~\figref{fig:fulldisp}{c}, loosely speaking, the number of periods is given by the number of times  how often a color reappears as one goes along the circumference of the hexagonal cross section. As an important difference to the cylindrical systems, that are characterized by full rotational symmetry about the tube axis, the phase around the hexagonal cross sections does not increase linearly along the circumference. In fact, the discrete sixfold rotational symmetry induces drastic qualitative changes. We observe that, depending on the number of periods, the  modes can be either doubly degenerate or non-degenerate, respectively forming duplets or singlets. 
%Specifically, when the number of periods along the circumference matches the symmetry of the hexagonal tube (0, 3, 6, 9, ...), for a given localization, only one solution (a singlet) exists, which forms a standing wave along the circumference. 
Singlets form standing waves along the circumference, as can be seen in the $\pi$ jumps of the phase of the C3 mode in Fig.~\figref{fig:fulldisp}{c}. 
%In contrast to this, for different period numbers (2, 4, 5, 7, 8,...) two degenerate solutions (duplet) exist, \textit{i.e.} in Fig.~\figref{fig:fulldisp}{a} every duplet branch consists of two overlapping ones. 
%As seen at the example of the branches $\textsc{c}2-$ and $\textsc{c}2+$ in Fig.~\figref{fig:fulldisp}{c}, the duplets consist of two solutions which have equal localization (magnitude) but are propagating in opposite directions along the circumference. 
In contrast to this, the duplets consist of two solutions which have equal localization (magnitude) but are propagating in opposite directions along the circumference, as depicted for C2+ and C2- in Fig. 3(c).
Albeit the phase does not increase in a linear fashion as in cylindrical systems, it still changes continuously. In Fig.~\figref{fig:fulldisp}{d,e}, we show the mode profiles close to $k=0$ for all branches whose dispersion is plotted in Fig.~\figref{fig:fulldisp}{a}. Notably, for all the modes the profile of the magnitude obey the sixfold rotational symmetry of the hexagonal tube. The observed splitting of the spectrum into singlets and duplets has also been observed in a similar way for whispering gallery modes in hexagonal optical cavities.\cite{yangWhisperinggalleryModeHexagonal2019}
%and can be discussed thoroughly in the context of group theory, more specifically representation theory. Let us note, however, that our system does not exhibit the same symmetry group as the regular hexagon ($D_6$) in Ref.~\citenum{yangWhisperinggalleryModeHexagonal2019}, since the magnetic vortex state breaks the mirror symmetries contained in $D_6$. Including the combined mirror and time reversal operations the magnetic point group of our system is  6/m'mm. This point group has only one and two dimensional irreducible representations, therefore the solutions should consist of singlets and duplets,\cite{IrreducibleCorepresentationsMagnetic} which have to be either standing or running waves along the hexagonal circumference, respectively.  \textbf{TODO for Istv\'an: Do you have a reference why singlets are standing while duplets are running waves?} 

\begin{figure}[h!]
    \centering
    \includegraphics[width=8.6cm]{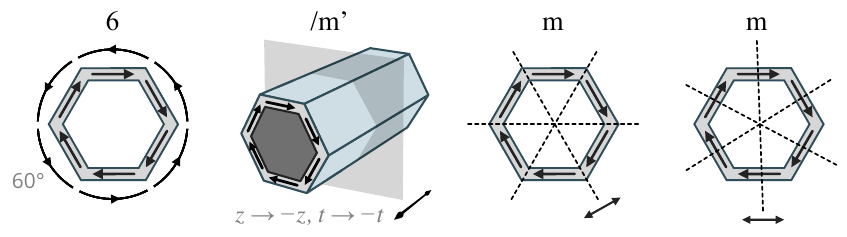}
    \caption{Generators of the symmetry group 6/m'mm shown for the hexagonal tube containing a vortex state.}
    \label{fig:generators} 
\end{figure}

In the following, we show how the splitting of some of the duplets to singlet pairs upon lowering the symmetry of the vortex tube from cylindrical to hexagonal can be understood via a basic group theory approach, even without considering the form of the magnetic interactions. The symmetry of the hexagonal vortex tube is described by the magnetic point group 6/m'mm. The generators of the group are shown in Fig.~\ref{fig:generators}: The 6-fold rotational symmetry about the axis of the tube, two sets of mirror planes (m) containing the axis of the tube and the mirror plane perpendicular to the tube axis (/m'). Due to the magnetic vortex pattern, the latter are only symmetries when combined with the time reversal operation ('). This magnetic point group has only one and two dimensional irreducible representations, thus the excitations of the hexagonal magnetic vortex can only form singlets and duplets. No modes with triple or higher degeneracy can emerge. The singlet solutions listed up in Fig.~\figref{fig:fulldisp}{d}, can be classified according to the one dimensional representations of 6/m'mm.\cite{IrreducibleCorepresentationsMagnetic} Since the amplitude map of all modes obey all symmetries in this group, one needs to check how the phase pattern of the modes change upon the different symmetry operations. For singlets, the phase pattern can either be invariant upon a symmetry operation or acquire a pi shift. Modes CF0, CF6 and C6 belongs to the fully symmetric $A_1$ irreducible representation, as their phase pattern is invariant upon all symmetry operations of the 6/m'mm group. The phase pattern of the F6 mode is invariant upon all symmetry operation but the reflections to the mirror planes containing the tube axis, thus it belongs to the $A_2$ irreducible representation. The classification is indicated for all the modes in Figs.~\figref{fig:fulldisp}{d,e}. %[[[Here is the full list for the figure: $A_1$: CF0, CF6, C6; $A_2$: F6; $B_1$: C3; $B_2$: F3, F9; $E_1$: CF4+/-, C5+/-, CF7+/-; $E_2$: C2+/-, F4+/-, CF8+/-]]] 
In case of duplets, some of the symmetry operations interrelate the phase patterns of the two modes, leading to a zero entry in the character table.\cite{IrreducibleCorepresentationsMagnetic}

One can easily find a correspondence between the modes of the cylindrical and the hexagonal vortex. The CF0 singlet corresponds to only singlet mode of the cylindrical tube, which is nothing but the simple ferromagnetic resonance for k=0. When the hexagonal symmetry is increased to cylindrical, the rest of the singlets become doubly degenerate, i.e. arrange into pairs, such as C6-F6, C3-F3, etc. This is because modes localized to corners and facets become non-distinguishable once the cylindrical symmetry is restored. 

We observe that modes with different periods around the circumference are hybridized, while it seems that singlets (duplets) only hybridize with singlets (duplets).  In Fig.~\figref{fig:fulldisp}{a}, the hybridization can be seen for the branches $\textsc{cf}5$ and $\textsc{cf}7$ and was confirmed by analyzing the mode profiles of the two branches on different sides of the crossings. To avoid visual clutter, we refrained from double labeling the branches twice. Let us also note that the dynamic-matrix approach used here can only yield the already hybridized normal modes of the system. Therefore, presenting a dispersion with the non-hybridized branches as well as a proper treatment of the hybridization would require an analytic theory which is not available at the moment. 

The same holds if one would like to disentangle the contributions resulting in the strongly asymmetric dispersion for some of the spin-wave modes. Based on our knowledge from the thin-shell cylindrical nanotubes we can state that the asymmetric dispersion has its origin in the  dynamic charges associated to the dipole-dipole interaction. Let us note, that the antiparallel alignment of the equilibrium magnetization in opposite facets would alone lead to an asymmetry, resulting in a linear shift of the dispersion in the small $k$ limit.\cite{gallardoReconfigurableSpinWaveNonreciprocity2019} However, in our case the asymmetry is far stronger than a linear shift and therefore suggests the presence of a geometrical volume charge due to the strongly curved regions between the flat facets. The concept of geometrical charges and its relation to possible magnetochiral effects is discussed in detail in Ref.~\citenum{shekaNonlocalChiralSymmetry2020}. The detailed discussion of the origin of the asymmetric dispersion as well as the presence of singlet and duplet states and their relation to the magnetic point group of our system is out of the purpose of the current manuscript and will be investigated in a forthcoming work.

\subsection{Predicted microwave absorption and comparison with experiments}

\begin{figure*}[t]
    \centering
    \includegraphics[width=14cm]{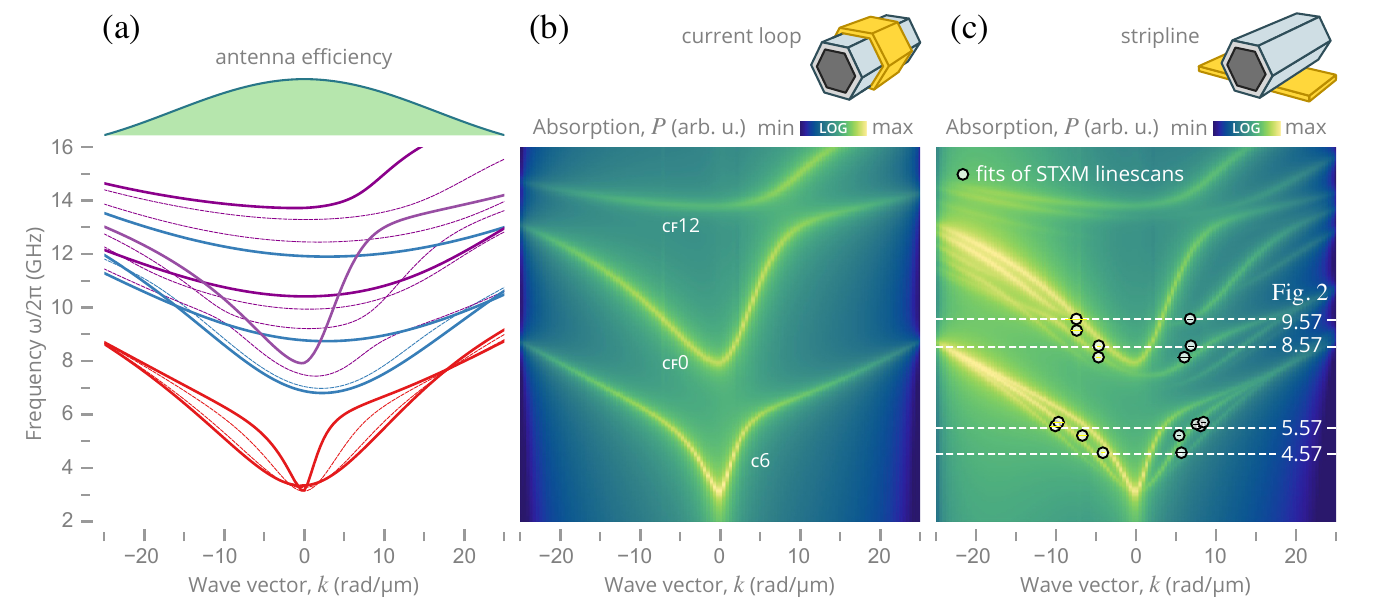}
    \caption{Predicted microwave absorption calculated using the mode profiles from the dynamic-matrix approach for two different antenna geometries. (a) The zeroth-order peak of the excitation efficiency at the surface of the antennas is shown on the top of the full dispersion relation including all modes up to \SI{16}{\giga\hertz}. The dispersion relation obtained when exciting spin-wave modes with a current loop antenna is shown in (b). Only a number of singlet branches are susceptible to the spatial distribution of such a field. The dispersion relation for a stip-line antenna used in the experiments is summarized in panel (c). The usual asymmetry of the excitation efficiency known for Damon-Eshbach modes is recovered. This antenna will excite simultaneously a multiple of modes with the same frequency. The white dashed lines indicate the excitation frequencies for which the experimental spatial spin-wave profiles are shown in Fig.~\ref{fig:fig2}.}
    \label{fig:absorption}
\end{figure*}

The dispersion branches excited in an experimental setup can drastically depend on the spatial distribution of the oscillating excitation field $\bm{h}(\bm{r})$, \textit{i.e.} on the microwave antenna at hand. In the case of propagating waves, the microwave power absorbed by the magnetic system is determined by the overlap 
\begin{equation}
    h_\nu (k) = \frac{1}{\mathcal{N}_\nu} \left\langle \bm{\eta}^*_{\nu k}\cdot \tilde{\bm{h}}(k) \right\rangle_A
\end{equation}
with $\tilde{\bm{h}}(k)$ denoting the Fourier transform of the spatial distribution of the microwave field with respect to the $z$ coordinate (long axis of the tube), $\langle...\rangle_A$ denotes the spatial average in the hexagonal-tube cross section and $\mathcal{N}_\nu$ is a normalization factor (see Appx.~\ref{appx:absorption}). The full frequency- and wave-vector-dependent absorption is then given by summing up the response of all branches $\nu$, as
\begin{equation}
    P(k,\omega) \propto \sum\limits_\nu \frac{\abs{h_\nu}^2(k)}{[\omega_\nu(k) - \omega]^2 - \Gamma_\nu^2(k)}.
\end{equation}
Here $\Gamma_\nu=\alpha_\mathrm{G} \epsilon_\nu \omega_\nu$ is the life time of the spin-wave modes which is determined by the Gilbert damping factor $\alpha_\mathrm{G}$ and by the mode ellipticity $\epsilon_\nu(k)$ (see Ref.~\citenum{verbaDampingLinearSpinwave2018} and again Appx.~\ref{appx:absorption})

In the following, we present the predicted absorption calculated for two different important antenna geometries, a single current loop wrapped around the nanotube and for a single stripline antenna attached to one of the facets, as the one used in our experiments. We fix both antennae width to $W = \SI{250}{\nano\meter}$. In Fig.~\figref{fig:absorption}{a}, on-top of the dispersion, we show the zeroth-order peak of the excitation efficiency at the surface of both antenna types, which is approximately given by $\abs{\mathrm{sinc}(\Lambda k)}$, with $\Lambda\approx W/2$. Exact expressions for the Fourier components of $\tilde{\bm{h}}(k)$ for the presented antennae are found in Appx.~\ref{appx:expressions}. As an important difference between the antenna geometries, the microwave field produced by a stripline antenna is inhomogeneous within the hexagonal cross section and is, therefore, not rotationally symmetric. While the microwave field produced by a current loop can be assumed to have a homogeneous magnitude along all facets. As a result, the spin-wave modes of particular symmetry propagating in the hexagonal tubes couple differently to the microwave excitation, depending on the specific field distribution. This knowledge is crucial when designing and interpreting experiments (and even classic time-domain micromagnetic simulations). To this end, in Fig.~\figref{fig:absorption}{b}, we show the absorption $P(k,\omega)$ for a current-loop microwave antenna. As can be seen, only a number of singlet branches (\textsc{c}6, \textsc{cf}0 and \textsc{cf}12) are susceptible to such a field. These are the only modes, which, according to their symmetry, exhibit a non-vanishing absorption in a rotiationally symmetric microwave field in the observed frequency range. The period of these modes is an integer multiple of 6 and in general the period of the singlets is $3n$, with $n \geq 0$.
%\textbf{TODO and they all have periods $6n$ periods}
%

In contrast to this, the stripline antenna used in our experiments will couple to the singlet as well as to the duplet modes, as shown in Fig.~\figref{fig:absorption}{c}. Note, that the excitation efficiency of the spin-wave modes is asymmetric, therefore for the considered vortex state and antenna geometry the spin waves propagating with negative wave vector are excited stronger than the counter propagating ones. As mentioned before, this is a commonly known effect for spin waves excited in the Damon-Eshbach geometry, namely $\bm{k}\perp\bm{m}_0$ and is also seen when analyzing the individual frames of the spin-wave spatial profiles obtained from the TR-STXM experiments. The white dashed lines mark the excitation frequencies for which the experimental and micromagnetic simulation spatial profiles of the spin waves are presented in Figure~\ref{fig:fig2}. The consequences of the stripline antenna microwave source are that multiple modes with the same frequency are excited simultaneously, leading to a beating pattern instead of a plane wave propagation pattern with well-defined wave length. The presence of multiple wave lengths can already be seen when carefully looking at the propagating spin-wave profiles taken at given snapshots in time from the time-domain micromagnetic simulations presented in Figure~\ref{fig:fig2}. Especially for the \SI{4.571}{\giga\hertz} and \SI{5.571}{\giga\hertz} frequencies it is quite obvious that multiple wavelengths are present. On one hand this explains why only certain frames from the experimental measurements could be used to approximately evaluate a wavelength for the excited spin waves. On the other hand, with this information in mind we need to emphasis that the determination of the wavelengths are rather imprecise. Still, without drawing conclusions, on Fig.~\figref{fig:absorption}{c} we have overlaid the experimentally determined dispersion with the one yielding from the predicted microwave absorption using the spin-wave mode profiles of the dynamic-matrix approach simulations. %As a positive outcome we can mention that we have directly measured spin-wave propagation in permalloy nanotubes with hexagonal cross section. 

\section{Conclusions and outlook}\label{sec:outlook}

We have investigated spin-wave propagation in hexagonal nanotubes using time-resolved STXM measurements and micromagnetic simulations. The experimental results show that spin waves can be excited with a simple stripline antenna. Using a finite element dynamic-matrix approach for propagating spin waves, we calculated the dispersion relation for the hexagonal tube with geometrical and material parameters as in the experiments. The dispersion relation turned out to be asymmetric and complex. Due to the hexagonal cross section, spin waves can be localized to the highly curved corners, to the flat facets as well as to both sites at the same time. The hexagonal symmetry lifts the azimuthal mode degeneracy known from round nanotubes and result in singlet and duplet spin-wave solutions. The singlets are always standing spin-wave solutions and their azimuthal mode index is an integer multiple of 3. The duplets consists of two degenerate spin-wave solutions counter propagating along the azimuthal direction. We have shown that, using the spin-wave profiles resulting from the eigensolver, the frequency- and wave-vector-dependent microwave absorption of different antennae field profiles can be calculated. These numerical results  show  that the stripline antenna used in the TR-STXM experiments will simultaneously excite modes with the same frequency but different wave vectors. Therefore, the resulting spatial profile of the spin waves propagating in the nanotube form a beating pattern instead of a single wave solution with a well-defined wavelength. These experimental results are not suitable to draw conclusions on the spin-wave dispersion asymmetry that is present in the dispersion calculated by micromagnetic simulations, both in the time- or frequency domain. We can conclude that the antenna design in further experiments needs to be changed and if possible a single current loop should be used to allow for the excitation of single modes with a well defined wave length. Alternatively one could use a CPW antenna which selectively excites spin waves with specific wave vectors. We hope that with the recent developments in materials research and fabrication methods the production of high quality 3D nano-structures and waveguides (magnetic for this purpose) will be standardized and the investigation of exciting effects as the curvature-induced magnetochiral effects on the magnetization statics and dynamics will become feasible.

\begin{acknowledgements}
The experiments were mainly performed at the MAXYMUS endstation of BESSY II at Helmholtz-Zentrum Berlin, Germany. We thank HZB for the allocation of synchrotron radiation beam time. Some experiments were performed at the PolLux endstation of the Swiss Light Source. We acknowledge the Paul Scherrer Institut, Villigen PSI, Switzerland for provision of synchrotron radiation beamtime. The PolLux end station was financed by the German Ministerium für Bildung und Forschung (BMBF) through contracts 05K16WED and 05K19WE2. Financial support by the Deutsche Forschungsgemeinschaft within the program KA 5069/1-1, KA 5069/3-1 and the project ID 422 314695032-SFB1277 is gratefully acknowledged. We also gratefully acknowledge financial support by the Fondecyt Iniciacion grant number 11190184. We thank M. Bechtel (MPI-IS) and B. Sarafimov (PSI) for technical support.

\end{acknowledgements}

\section*{Appendix}
\begin{appendix}

\section{Microwave absorption and linewidths}\label{appx:absorption}

Here, we briefly describe how the microwave absorption is calculated from the lateral mode profiles $\bm{\eta}_{\nu k}$ obtained with our propagating-wave dynamic-matrix approach. As mentioned in the main text, the microwave power absorbed by the spin-wave system is determined by the overlap of the spin-wave mode profile with the spatial profile of the microwave field. Our formalism here is a special case of the general cases \textit{e.g.} discussed in Refs. \citenum{naletovIdentificationSelectionRules2011} and \citenum{verbaDampingLinearSpinwave2018}. For a general volumentric spin-wave mode profile $\bm{m}_\nu(\bm{r})$ denoted only with the mode index $\nu$, the microwave absorption is obtained as 
\begin{equation}\label{eq:3doverlap}
    h_\nu = \frac{1}{\mathcal{M}_{\nu} V}\int\limits_V \mathrm{d}V^\prime \, \bm{m}^*_{\nu}(\bm{r}^\prime)\cdot {\bm{h}}(\bm{r}^\prime)
\end{equation}
with $V$ being the volume of the magnetic specimen, $\bm{h}$ being the spatial profile of the microwave field and $\mathcal{M}_\nu$ being the normalization factor of the mode with respect the volume, which is given by
\begin{equation}\label{eq:3dnormalizationfactor}
    \mathcal{M}_\nu = \frac{i}{V}\int\limits_V \mathrm{d}V^\prime \, \bm{m}^*_{\nu}(\bm{r}^\prime)\cdot [\bm{m}_0(\bm{r})^\prime\times \bm{m}_{\nu}(\bm{r}^\prime)].
\end{equation}
In our case, the mode profiles are given as $\bm{m}_\nu (\bm{r}) = \bm{\eta}_{\nu k} \exp(ikz)$ and the equilibrium magnetization is translationally invariant along the $z$ direction, $\bm{m}_0(\bm{r}) = \bm{m}_0(\bm{\rho})$. For a very long waveguide with finite length $L$, one can now insert these mode profiles into Eqs.~\eqref{eq:3doverlap} and \eqref{eq:3dnormalizationfactor}, perform the integral along the $z$-direction and then let $L\rightarrow\infty$. One then obtains the wave-vector dependent overlap
\begin{equation}
    h_\nu (k) = \frac{1}{\mathcal{N}_{\nu k} A}\int\limits_A \mathrm{d}A^\prime \, \bm{\eta}^*_{\nu k}(\bm{\rho}^\prime)\cdot \tilde{\bm{h}}(\bm{\rho^\prime},k)
\end{equation}
with $\tilde{\bm{h}}(\bm{\rho},k)$ being the Fourier transform of the microwave field along the $z$ direction and $\mathcal{N}_{\nu k} = \mathcal{M}_{\nu k}/L$ being the normalization factor of the mode with respect the cross section area $A$, which is given as
\begin{equation}
    \mathcal{N}_{\nu k} = \frac{1}{ A}\int\limits_A \mathrm{d}A^\prime \, \bm{\eta}^*_{\nu k} (\bm{\rho}^\prime) \cdot [\bm{m}_0(\bm{\rho}^\prime) \times  \bm{\eta}_{\nu k} (\bm{\rho}^\prime)].
\end{equation}
In order to obtain the full microwave absorption, one also needs to know the linewidths of the modes, $\Gamma_{\nu k }= \alpha_\mathrm{G}\epsilon_{\nu k}\omega_{\nu k} $, which depend on the Gilbert damping parameter $\alpha_\mathrm{G}$ and the mode ellipticity $\epsilon_{\nu k}$. A general formalism to obtain the linear spin-wave damping from the mode ellipticities was presented in Ref.~\citenum{verbaDampingLinearSpinwave2018}, which can be applied to our case in the same way as above. For our case we obtain,
\begin{equation}
\begin{split}
        \epsilon_{\nu k} & =  \frac{1}{\mathcal{N}_{\nu k} A}\int\limits_A \mathrm{d}A^\prime \,\abs{ \bm{\eta}_\nu(\bm{\rho}^\prime)}^2.
\end{split}
\end{equation}

\section{Expressions for microwave-field wave-vector spectra}\label{appx:expressions}

In our specific study, we calculated the microwave absorption for a current-loop antenna around and a stripline antenna attached to the hexagonal nanotube. In the case of a current-loop the wave-vector spectrum of the microwave field can be approximated as 
\begin{equation}
    \tilde{\bm{h}}^{\mathrm{(loop)}} \approx (\bm{e}_\xi(\bm{\rho}) + \bm{e}_z) \abs{\mathrm{sinc}(\Lambda k)} e^{-\beta_0 k}
\end{equation}
with $\Lambda \approx \SI{125}{\nano\meter}$ being approximately equal to half of the width of the antenna and $\beta_0 = \SI{0.5}{\nano\meter/\radian}$ being some decay factor. The unit vector field $\bm{e}_\xi $ can be approximated as being the one locally perpendicular to both $\bm{m}_0$ and $\bm{e}_z$, which, in the case of a hexagonal vortex state, gives the "radial" direction.

In the case of a stripline antenna which is attached to on the facets of the hexagonal tube (w.l.o.g. a facet which is parallel to the $xz$ plane) the wave-vector spectrum can be approximated as
\begin{equation}
    \tilde{\bm{h}}^{\mathrm{(strip)}} \approx (\bm{e}_x + \bm{e}_z) \abs{\mathrm{sinc}(\Lambda k)} e^{-\beta(s)k}.
\end{equation}
Here, $\beta(s)$ is now function which depends on the distance $s$ to the center plane of the antenna. To obtain this dependence we calculated the full 3D profile of the stripline antenna and performed a Fourier transform in $z$ direction for different distances from the antenna. A simple linear approach gave a reasonable fit $\beta(s) = \beta_1 s + \beta_0$ with $\beta_1 \approx 1 $ and again $\beta_0 = \SI{0.5}{\nano\meter/\radian}$.
\end{appendix}
%\begin{suppinfo}

%The following files are available free of %charge.
%\begin{itemize}
%\item PDF describing the methods used in this paper, the indentification of the higher-order top-bottom modes as well as the influence of oppositely magnetized facets on the dispersion asymmetry.
%  \item movie\_4p571GHz.gif: Time-resolved STXM measurement at 4.571 GHz excitation frequency. Field of view: 2um x 0.5um in 100pt x 25pt
%  \item movie\_5p571GHz.gif: Time-resolved STXM measurement at 5.571 GHz excitation frequency. Field of view: 3um x 0.5um in 150pt x 25pt
%    \item movie\_8p571GHz.gif: Time-resolved STXM measurement at 8.571 GHz excitation frequency. Field of view: 4um x 0.5um in 200pt x 25pt
%  \item movie\_9p571GHz.gif: Time-resolved STXM measurement at 9.571 GHz excitation frequency. Field of view: 3um x 0.5um in 150pt x 25pt
%\end{itemize}

%\end{suppinfo}

\bibliographystyle{apsrev4-1}

%\bibliography{references}

\end{document}